\documentclass[a4paper,11pt]{article}
\pdfoutput=1 

\usepackage{jcappub} 

\usepackage[T1]{fontenc} 
\usepackage{subcaption}

\title{A generalized method of constraining Warm Inflation with CMB data}

\author[a,1]{Umang Kumar\note{Corresponding author}}
\author[a]{and Suratna Das}

\affiliation[a]{Department of Physics, Ashoka University, Rajiv Gandhi Education City, Rai, Sonipat: 131029, Haryana, India}

\emailAdd{umang.kumar$\_{}$phd21@ashoka.edu.in}
\emailAdd{suratna.das@ashoka.edu.in}

\abstract{
A thorough MCMC analysis of any inflationary model against the current cosmological data is essential for assessing the validity of such a model as a viable inflationary model. Warm Inflation, producing both thermal and quantum fluctuations, yield a complex form of scalar power spectrum, which, in general, cannot be directly written as a function of the comoving wavenumber $k$, an essential step to incorporate the primordial spectra into CAMB to do an MCMC analysis through CosmoMC/Cobaya. In this paper, we devised an efficient generalized methodology to mould the WI power spectra as a function of $k$, without the need of slow-roll approximation of the inflationary dynamics. The methodology is directly applicable to any Warm Inflation model, including the ones with complex forms of the dissipative coefficient and the inflaton potential. 
}

\begin{document}
\maketitle
\flushbottom

\section{Introduction}
\label{intro}

The full-mission Planck measurements of the Cosmic Microwave Background anisotropies   \cite{Planck:2018vyg} have indeed constrained the standard spatially-flat 6-parameter $\Lambda$CDM cosmology with a power-law spectrum of adiabatic scalar perturbations very well. The six base parameters of Planck observations \cite{Planck:2018vyg} are $\Omega_bh^2$ (baryon density parameter), $\Omega_ch^2$ (Dark Matter density parameter), $100\theta_{\rm MC}$ (acoustic angular scale), $\tau$ (ionization optical depth), $\ln(10^{10} A_s)$ (scalar density perturbation amplitude $A_s$) and $n_s$ (scalar density perturbation spectral index). $H_0$ (the current Hubble constant), on the other hand, is one of the derived parameters of Planck observation. These parameters are measured by the Planck observation (TT+TE+EE+lowE+lensing) as follows \cite{Planck:2018vyg}:
\begin{eqnarray}
&&\Omega_bh^2=0.02237\pm0.00015, \quad\quad \Omega_ch^2=0.1200\pm0.0012, \quad\quad 100\theta_{\rm MC}=1.04092\pm0.00031,\nonumber\\
&&\tau=0.0544\pm0.0073, \quad\quad \ln(10^{10}A_s)=3.044\pm0.014, \quad\quad n_s=0.9649\pm0.0042,\nonumber\\
&& H_0[{\rm km\,s^{-1}\,Mpc^{-1}}]=67.36\pm0.54.\nonumber
\end{eqnarray}
Moreover, combining with the BICEP2/Keck Array BK15 data \cite{BICEP2:2015nss}, Planck has put a stringent bound on the ratio, $r$, of the amplitudes of the tensor and scalar density perturbations as  \cite{Planck:2018jri}
\begin{eqnarray}
r_{0.002}<0.056.\nonumber
\end{eqnarray}
Here $0.002$ (Mpc$^{-1}$) signifies the pivot scale at which $r$ has been constrained. As only two primordial parameters, $A_s$ and $n_s$, are measured by Planck in addition to putting an upper bound on $r$, there are still a plethora of inflationary models which can satisfy the current data \cite{Martin:2013tda}. On the other hand, these parameters help constrain the model parameters of any viable inflationary model. Thus, it is essential to put to test any inflationary model against the current cosmological data to assess its validity. For this purpose, there are publicly available codes, like CosmoMC \cite{Lewis:2002ah, Lewis:2013hha} or Cobaya \cite{Torrado:2020dgo}, through which one can perform Markov-Chain Monte Carlo (MCMC) analysis of the inflationary model parameters. Both these codes use CAMB  \cite{Lewis:1999bs, Howlett:2012mh}, another publicly available code, to incorporate the primordial power spectra, both scalar and tensor, as inputs. 

Warm inflation (WI) \cite{Berera:1995ie} is an alternate inflationary scenario where the inflaton field keeps dissipating its energy density to a nearly constant radiation bath during inflation, and as a result, WI smoothly transitions into a radiation dominated epoch without invoking a requirement of a reheating phase post inflation (for more recent reviews on WI see \cite{Kamali:2023lzq, Berera:2023liv}). In the standard (cold) inflationary scenario (see lecture notes on cold inflation \cite{Riotto:2002yw, Baumann:2009ds, Mishra:2024axb}), the coupling of the inflaton field with other degrees of freedom are usually neglected, and any energy density present before inflation starts, dilutes away exponentially fast during inflation, ending in a cold Universe which is then required to be  subsequently reheated to enter a standard radiation dominated epoch. The actual mechanism of reheating is largely unknown, and a reheating phase doesn't, in general, leave any observational signature \cite{Riotto:2002yw, Baumann:2009ds, Mishra:2024axb}. WI, by construction, overcomes the requirement of a post-inflationary reheating phase, and thus devoid of any ambiguity in the evolutionary history of our Universe that a reheating phase in general brings in. 

WI has certain advantages over the standard cold inflationary scenario. Due to the presence of the subdominant radiation bath during WI, both quantum and thermal (classical) primordial perturbations are produced during WI, and the thermal fluctuations dominate the scalar primordial power spectrum \cite{Ramos:2013nsa, Hall:2003zp, Graham:2009bf}. 
Due to this feature, on one hand, WI doesn't face the problem of quantum-to-classical transition of primordial perturbations, which is another daunting issue with the standard cold inflation \cite{Polarski:1995jg, Kiefer:2008ku, Kiefer:2006je, Kiefer:1998qe, Martin:2012pea, Das:2013qwa, Das:2014ada}, and on the other hand, reduces the tensor-to-scalar ratio $r$ \cite{Ramos:2013nsa, Hall:2003zp, Graham:2009bf}. Because of this second feature, WI can accommodate simple chaotic monomial potentials like, the quadratic \cite{Bastero-Gil:2019gao} and the quartic \cite{Bartrum:2013fia}, which are otherwise observationally ruled out in cold inflation for producing large $r$ \cite{Planck:2018jri}. Certain WI scenarios produce Primordial Black Holes (PBHs) naturally \cite{Arya:2019wck, Bastero-Gil:2021fac, Correa:2022ngq} without requiring deviation from slow-roll dynamics (like a transient ultraslow-roll phase \cite{Motohashi:2017kbs} or a bump/dip in the potential \cite{Mishra:2019pzq}). Currently, PBHs are considered as a favoured candidate for Dark Matter \cite{Carr:2016drx}. Moreover, due to its dissipative effect, WI introduces extra frictional term in inflaton's dynamics, which help the inflaton field slow-roll even along very steep potentials. Thus WI can accommodate steep potentials \cite{Das:2020xmh} whereas such potentials are unsuitable for standard cold inflation scenario. Due to this very feature, WI can naturally overcome the de Sitter Swampland conjecture \cite{Das:2018hqy, Motaharfar:2018zyb, Das:2018rpg, Das:2019hto} recently proposed in String Theory \cite{Obied:2018sgi, Garg:2018reu}, and has become a more natural inflationary scenario, than the cold version, for UV-complete gravity theories.     

However, due to the thermal primordial fluctuations, the form of the scalar power spectrum in WI differs significantly from the standard cold inflation. A few attempts have been made previously \cite{Benetti:2016jhf, Arya:2017zlb, Bastero-Gil:2017wwl} to incorporate such non-trivial WI power spectrum into CAMB (and subsequently into CosmoMC) in order to perform a MCMC analysis of the model parameters. Yet, the prescriptions formulated previously in \cite{Arya:2017zlb, Bastero-Gil:2017wwl} are quite restrictive, and are only applicable to very specific WI models with rather simple forms of inflaton potentials. Such prescriptions fail to incorporate more intricate WI models into CAMB, and hence it is not possible to put to test non-trivial WI models against the current data with the help of those existing prescriptions. In this paper, we have devised a generalized methodology to incorporate any WI power spectrum into CAMB much more efficiently than the existing methodologies. It is to note that recently a new numerical method based on Fokker-Planck approach (referred to as matrix formalism) has been developed to calculate the scalar power spectrum in WI \cite{Ballesteros:2023dno}, which doesn't yield any analytical form of the power spectrum to be fed into CAMB. Hence, a MCMC analysis of the WI scalar power spectrum obtained from such an approach cannot be performed, and the parameters can only be constrained by scanning the whole parameter space \cite{Ballesteros:2023dno}. 

We have furnished the rest of the paper as follows. In Sec.~\ref{wi-brief} we have briefly discussed with generic features of any WI model. In Sec.~\ref{limitations}, we have commented on the limitations of previous methodologies which then justifies the need of a more generalized methodology to be developed. In Sec.~\ref{method}, we have provided a step-by-step generalized methodology to incorporate any WI power spectra into CAMB which will then help analyse any such model with the current data (with the MCMC analysis done by CosmoMC or Cobaya). In Sec.~\ref{tests}, we have first analysed one Warm Inflationary scenario which couldn't have been analysed by the previous methodologies. This signifies the advantages of this newly developed methodology over the previous ones. Then we reanalysed two WI models which were previously studied in using the previous methodologies, and showed that our method can produce the previous results quite well. This validates the functionality of our newly developed methodology. At the end, in Sec.~\ref{conclusion}, we conclude.

\section{A brief discussion on generic features of Warm Inflation}
\label{wi-brief}

The inflaton field, $\phi$, dissipates its energy to a nearly constant subdominant radiation bath during a Warm Inflationary scenario. Therefore, there are two kinds of energy densities, inflaton energy density (both kinetic energy density and potential energy density) and the radiation energy density $\rho_r$,  present in the universe during WI. The evolution equations of both these entities,  
\begin{eqnarray}
&&\ddot\phi+3H\dot\phi+V,_\phi=-\Upsilon\dot\phi, \label{KG-eq} \\
&&\dot\rho_r+4H\rho_r=\Upsilon\dot\phi^2, \label{rhor-eq}
\end{eqnarray}
along with the Friedmann equations,
\begin{eqnarray}
3M_{\rm Pl}^2H^2&=&\frac12\dot\phi^2+V(\phi)+\rho_r, \label{first-friedmann}\\
-2M_{\rm Pl}^2\dot H&=&\dot\phi^2+\frac43\rho_r, \label{second-friedmann}
\end{eqnarray}
set the background dynamics of WI. In the above equations, overdot represents derivative with respect to the cosmic time $(t)$ and $M_{\rm Pl}$ is the reduced Planck mass. In Eq.~(\ref{KG-eq}) and Eq.~(\ref{rhor-eq}), $\Upsilon$ designates the dissipative coefficient of the inflaton field which determines the rate at which the inflaton energy density transfers to the radiation energy density. Any generic WI scenario is analysed assuming a near thermal equilibrium of the radiation bath, due to which one can associate a temperature $T$ with the radiation bath, and one can write 
\begin{eqnarray}
\rho_r=\frac{\pi^2}{30}g_*T^4\equiv C_RT^4, \label{rho-T}
\end{eqnarray}
where $g_*$ is the relativistic degrees of freedom present in the thermal bath. Many WI models have been studied in such a near thermal equilibrium condition, and it appears that the dissipative coefficients of such models take a generic form:
\begin{eqnarray}
\Upsilon(\phi,T)=C_\Upsilon T^p\phi^cM^{1-p-c}, \label{ups}
\end{eqnarray}
where $C_\Upsilon$ is a dimensionless quantity that depends on the underlying microphysics of the WI model under consideration, and $M$ is an appropriate mass scale so that the dimension of the dissipative coefficient remains as [Mass] (in natural units). 

One can see from Eq.~(\ref{KG-eq}) that the inflaton's equation of motion has two friction terms in WI, one due to the Hubble expansion $(3H\dot\phi)$, which is also present in CI, and the other due to inflaton's dissipation $(\Upsilon\dot\phi)$. Depending on which one of these two frictional terms dominates the dynamics of the inflaton field, one can classify WI models into two regimes. For that, it is convenient to define the following dimensionless quantity:
\begin{eqnarray}
Q=\frac{\Upsilon}{3H},
\end{eqnarray}
which is nothing but the ratio of the two frictional terms present in the inflaton's equation of motion. The regime when $Q\ll1$, i.e., when the dynamics of the inflaton is dominated by the Hubble friction, much like in the standard CI, is called the weak dissipative regime. One of the well-know dissipative terms that leads to observationally viable weak dissipative WI is given as \cite{Berera:2008ar, Bastero-Gil:2010dgy, Bastero-Gil:2012akf}:
\begin{eqnarray}
\Upsilon_{\rm cubic}=C_\Upsilon \frac{T^3}{\phi^2}. \label{cubic}
\end{eqnarray}
The other noteworthy WI model which has been put to test in the weak dissipative regime is the Warm Little Inflaton model \cite{Bastero-Gil:2016qru}, where the dissipative coefficient takes the form:
\begin{eqnarray}
\Upsilon_{\rm linear}=C_\Upsilon T.  \label{linear}
\end{eqnarray}
On the other hand, the regime where $Q\gg1$, i.e. the dissipative term dominates over the Hubble friction in inflaton's equation of motion, is known as strong dissipative regime. The recently proposed Minimal Warm Inflation (MWI) model \cite{Berghaus:2019whh} successfully realises WI in the strong dissipative regime, and the dissipative coefficient of this model has a form:
\begin{eqnarray}
\Upsilon_{\rm MWI}=C_\Upsilon\frac{T^3}{M^2}. \label{MWI}
\end{eqnarray}
Another WI model proposed in \cite{Bastero-Gil:2019gao} with a dissipative term 
\begin{eqnarray}
\Upsilon_{\rm EFT}=C\frac{\tilde M^2T^2}{m^3(T)}\left[1+\frac{1}{\sqrt{2\pi}}\left(\frac{m(T)}{T}\right)^{3/2}\right]e^{-m(T)/T}, \label{EFT}
\end{eqnarray}
is also constructed to be realised in the strong dissipative regime.  Here $m^2(T)=m_0^2+\alpha^2T^2$ is the thermal mass for the light scalars coupled to the inflaton field in this model, with $m_0$ being the vacuum mass of these light scalars and $\alpha$ being the coupling constant. When the effective mass is dominated by its thermal part, $m(T)\sim\alpha T$, the dissipative coefficient becomes an inverse function of $T$ $(\Upsilon_{\rm EFT}\sim C_\Upsilon M^2/T)$. On the other hand, when the temperature of the radiation bath falls below the vacuum term $m_0$, the behaviour of the dissipative term changes and in the limiting case it becomes proportional to $T^2$ $(\Upsilon_{\rm EFT}\sim C_\Upsilon T^2/M)$.

During slow-roll of the inflaton field, one can ignore the $\ddot\phi$ term in Eq.~(\ref{KG-eq}). In such a case, the approximated equation of motion of the inflaton field becomes 
\begin{eqnarray}
3H(1+Q)\dot\phi\simeq -V,_\phi.
\end{eqnarray}
During inflation, the potential energy density of the inflaton field dominates, and the the first Friedmann equation, given in Eq.~(\ref{first-friedmann}), can be approximated as 
\begin{eqnarray}
3M_{\rm Pl}^2H^2\simeq V(\phi).
\end{eqnarray}
Furthermore, during WI, it is assumed that a nearly constant radiation bath is maintained throughout. Thus, Eq.~(\ref{rhor-eq}) can be approximated as 
\begin{eqnarray}
\rho_r\simeq \frac34 Q\dot\phi^2. 
\end{eqnarray}
Using this above equation, one can write the second Friedmann equation, given in Eq.~(\ref{second-friedmann}) as 
\begin{eqnarray}
-2M_{\rm Pl}^2\dot H\simeq (1+Q)\dot\phi^2.
\end{eqnarray}
Therefore, during slow-roll WI the Hubble slow-roll parameter $\epsilon_H$ can be determined as 
\begin{eqnarray}
\epsilon_H\equiv -\frac{\dot H}{H^2}\simeq\frac{M_{\rm Pl}^2}{2(1+Q)}\left(\frac{V,_\phi}{V}\right)^2=\frac{\epsilon_V}{1+Q}, \label{eH}
\end{eqnarray}
where we have defined the standard potential slow-roll parameter as $\epsilon_V=(M_{\rm Pl}^2/2)(V,_\phi/V)^2$. This reflects an extremely interesting feature of WI. It signifies that when WI takes place in strong dissipative regime ($Q\gg1$), one can accommodate very steep potentials (for which $\epsilon_V\gg1$) in WI while satisfying the slow-roll condition $\epsilon_H\ll1$. Such potentials are not suitable for CI as slow-roll doesn't take place with such steep potentials. In literature, steep potentials have been successfully accommodated in the strongly dissipative MWI model \cite{Das:2019acf, Das:2020xmh}. In tune with the above discussion, we define the potential slow-roll parameters in WI as follows:
\begin{eqnarray}
\epsilon_{\rm WI}\equiv \frac{\epsilon_V}{1+Q}=\frac{M_{\rm Pl}^2}{2(1+Q)}\left(\frac{V,_\phi}{V}\right)^2, \quad\quad \eta_{\rm WI}\equiv\frac{\eta_V}{1+Q}=\frac{M_{\rm Pl}^2}{1+Q}\frac{V,_{\phi\phi}}{V}.
\end{eqnarray}
We also note that  graceful exit of WI happens when $\epsilon_H\sim1$, or when 
$\epsilon_V\sim (1+Q)$.

The scalar power spectrum generated during WI significantly differs from the one generated during CI, solely due to the fact that WI generates both quantum as well as thermal (classical) fluctuations. The scalar perturbations and its spectrum generated during WI have been analyzed in the literature \cite{Ramos:2013nsa, Hall:2003zp, Graham:2009bf, Bastero-Gil:2011rva}, and the scalar power spectrum in WI can be written in the form:
\begin{eqnarray}
{\mathcal P}_{\mathcal R}(k_*)=\left(\frac{H_*^2}{2\pi\dot\phi_*}\right)^2\left(1+2n_*+\frac{2\sqrt{3}\pi Q_*}{\sqrt{3+4\pi Q_*}}\frac{T_*}{H_*}\right)G(Q_*). 
\label{scalar-ps}
\end{eqnarray}
The subindex $``*"$ indicates that those quantities are evaluated at horizon crossing ($k=aH$). Moreover, $n_*$ signifies the thermal distribution of the inflaton field in scenarios where it thermalizes with the radiation bath. It is customary to take $n_*=0$ if the inflaton field doesn't thermalize with the radiation bath. Otherwise, it is natural to assume an equilibrium Bose-Einstein distribution of the inflaton field as 
\begin{eqnarray}
n_*=\frac{1}{\exp\left(\frac{H_*}{T_*}\right)-1}.
\end{eqnarray}
However, it has been shown in \cite{Bastero-Gil:2017yzb} that as the temperature of the radiation bath and the Hubble parameter evolve slowly during WI, the inflaton field evolves in an out-of-equilibrium state (which they refer to as an adiabatic state) resulting in an $n_*$ between 0 and the equilibrium Bose-Einstein distribution. However, for a large decay width of the inflaton field, the distribution will soon reach the equilibrium state, and the distribution can be given as the standard Bose-Einstein distribution \cite{Bastero-Gil:2017yzb}. 
The $G(Q_*)$ factor in Eq.~(\ref{scalar-ps}) appears in the power spectrum as a signature of the coupling of the inflaton field with the radiation bath, and this growth factor can only be evaluated by numerically solving the full set of perturbation equations of WI \cite{Graham:2009bf, Bastero-Gil:2011rva, Bastero-Gil:2014jsa}. The functional form of $G(Q_*)$ primarily depends on the form of the dissipative coefficient and has only weak dependence on the form of the potential. The $G(Q_*)$ factor for the dissipative coefficient given in Eq.~(\ref{cubic}) can be given as  \cite{Benetti:2016jhf, Arya:2017zlb}
\begin{eqnarray}
G_{\rm cubic}(Q_*)\simeq 1+4.981Q_*^{1.946}+0.127Q_*^{4.330},
\end{eqnarray}
where as for the one given in Eq.~(\ref{linear}) one gets \cite{Benetti:2016jhf, Bastero-Gil:2017wwl}
\begin{eqnarray}
G_{\rm linear}(Q_*)\simeq 1+0.335Q_*^{1.364}+0.0185Q_*^{2.315}.
\end{eqnarray}
As the dissipative coefficient given in Eq.~(\ref{cubic}) has a higher temperature dependence on $T$ than the one in Eq.~(\ref{linear}), which implies stronger coupling between the inflaton and the radiation bath in the former case, the growth factor $G(Q_*)$ for the former has a higher $Q$ dependence than the later. The growth factor for the MWI dissipative coefficient, given in Eq.~(\ref{MWI}), was determined in \cite{Das:2020xmh} as 
\begin{eqnarray}
G_{\rm MWI}(Q_*)=\frac{1+6.12Q_*^{2.73}}{(1+6.96Q_*^{0.78})^{0.72}}+\frac{0.01Q_*^{4.61}(1+4.82\times 10^{-6}Q_*^{3.12})}{(1+6.83\times 10^{-13}Q_*^{4.12})^2},
\end{eqnarray}
whereas the growth factor for the one given in Eq.~(\ref{EFT}) was not explicitly spelt out in \cite{Bastero-Gil:2019gao}. Recently, a publicly available code, WarmSPy, has been developed in \cite{Montefalcone:2023pvh} to solve for the factor $G(Q)$ in any WI model. The Planck observation has set the value of the scalar spectral amplitude at $2.1\times 10^{-9}$ as the pivot scale $k_P=0.05$ Mpc$^{-1}$. 

In WI, the tensors modes doesn't get affected by the presence of the radiation bath as there are no direct couplings between them, and thus, WI yields the same tensor power spectrum as in CI:
\begin{eqnarray}
{\mathcal P}_T(k_*)=\frac{H_*^2}{2\pi^2M_{\rm Pl}^2}, \label{tensor-ps}
\end{eqnarray}
and the tensor-to-scalar ratio $r$ in WI can be determined as $r={\mathcal P}_T/{\mathcal P}_{\mathcal R}$.

\section{Limitations of former approaches in incorporating the WI scalar power spectrum in CAMB (and subsequently in CosmoMC/COBAYA) in general}
\label{limitations}

To constrain the model parameters of any inflationary model, one needs to perform MCMC analysis of these parameters against the most recent precision data. The MCMC analysis of inflationary model parameters can be done using the publicly available code CosmoMC \cite{Lewis:2002ah, Lewis:2013hha} or Cobaya \cite{Torrado:2020dgo} (with fast dragging procedure \cite{Neal:2005uqf}) which is a Python based code adapted from CosmoMC. Both these codes use CAMB \cite{Lewis:1999bs, Howlett:2012mh} to help incorporate the inflationary power spectra as an input.\footnote{Another way to incorporate the primordial power spectra in CosmoMC/Cobaya is through another publicly available code CLASS \cite{Lesgourgues:2011re}.} However, the only way the primordial power spectra can be fed into CAMB is by expressing them as a functions of the comoving wavenumber $k$ of the primordial perturbations. Thus to feed in the Warm Inflationary scalar and tensor power spectra, as given in Eq.~(\ref{scalar-ps}) and Eq.~(\ref{tensor-ps}) respectively, into CAMB, one needs to express them as functions of $k$.

The first attempt to put WI to test with (Planck) observations was made in \cite{Benetti:2016jhf}, where WI models with both the cubic (Eq.~(\ref{cubic})) and linear (Eq.~(\ref{linear})) dissipative coefficients were analysed combining with five different inflaton potentials: quartic, sextic, hilltop, Higgs and plateau sextic. However, this work doesn't shed any light on how one can write the primordial scalar power spectrum of WI, given in Eq.~(\ref{scalar-ps}), as a function of $k$, an essential step to incorporate the power spectrum into CosmoMC through CAMB. Soon after, two papers \cite{Bastero-Gil:2017wwl, Arya:2017zlb} attempted to test simple WI models with the Planck data. The first paper \cite{Bastero-Gil:2017wwl} took a WI model with the linear dissipative coefficient, given in Eq.~(\ref{linear}) and chose simple quartic potential ($\lambda \phi^4$) for the inflaton field. The other work analysed a WI model with the cubic dissipative term (Eq.~(\ref{cubic})) but again with the simple quartic inflaton potential. Both these works adopted similar methodologies to express the primordial power spectra (both scalar and tensor) in terms of $k$. We briefly describe the methodology adopted by these two works below.

The main step in expressing the scalar power spectrum in terms of $k$, is to first express it in terms of $Q$ alone. The reason behind this step is that, with a general dissipative coefficient, given in Eq.(\ref{ups}), one can estimate how $Q$ evolves with the $e-$foldings $N$ \cite{Das:2020lut} (assuming slow-roll and existence of a near-constant radiation bath), which can be written as follows:
\begin{eqnarray}
C_Q\frac{d\ln Q}{dN}=(2p+4)\epsilon_V-2p\eta_V-4c\kappa_V, \label{dQdN}
\end{eqnarray}
where $C_Q\equiv4-p+(4+p)Q$, which is always positive, and $\kappa_V\equiv M_{\rm Pl}^2(V,_\phi/\phi V)$. On the other hand, one can determine $N$ as a function of $k$, which, in turn, helps determine $Q$ as a function of $k$. Noting that $dN=d\ln a=Hdt$  and $k=aH$ at horizon crossing, one can write \cite{Das:2022ubr}
\begin{eqnarray}
\frac{d\ln(k/k_P)}{dN}=1-\epsilon_H, \label{dkdN}
\end{eqnarray}
where $k_P$ is the pivot scale, and $\epsilon_H$ is defined in Eq.~(\ref{eH}).\footnote{Note that in \cite{Bastero-Gil:2017wwl, Arya:2017zlb} the authors have counted the number of $e-$folds backward, such that there $N=0$ designates end of inflation. In our prescription $N=0$ designates the beginning of inflation.}

Looking at Eq.~(\ref{scalar-ps}), we see that there are two factors $H^2/\dot\phi$ and $T/H$ (which also helps express $n_*$ in terms of $Q$) which are required to be expressed in terms of $Q$. For the quartic $\lambda\phi^4$ potential the first term $H^2/\dot\phi$ can be written as 
\begin{eqnarray}
\frac{H^2}{\dot\phi}=\frac{1+Q}{\sqrt3 M_{\rm Pl}^3}\frac{V^{3/2}}{V,_\phi}=\frac{(1+Q)\sqrt\lambda}{4\sqrt3}\left(\frac{\phi}{M_{\rm Pl}}\right)^3,
\end{eqnarray}
which makes it a function of $Q$ as well as $\phi$. For the linear dissipative coefficient $T/H$ can be expressed as \cite{Bastero-Gil:2017wwl}
\begin{eqnarray}
\left(\frac TH\right)_{\rm linear}=\frac{3Q}{C_\Upsilon},
\end{eqnarray}
which makes it a function of $Q$ alone, whereas for the cubic dissipative term one gets \cite{Arya:2017zlb}
\begin{eqnarray}
\left(\frac{T}{H}\right)_{\rm cubic}=\left(\frac{1080}{\pi^2 g_*\lambda}\right)^{1/4}\frac{Q^{1/4}}{(1+Q)^{1/2}}\left(\frac{\phi}{M_{\rm Pl}}\right)^{-3/2},
\end{eqnarray}
which makes it a function of both $Q$ and $\phi$.\footnote{$M_{\rm Pl}$ in \cite{Arya:2017zlb} is defined as the Planck mass, not as the reduced Planck mass.} Therefore, the job will be done if one can express $\phi$ in terms of $Q$. Under slow-roll conditions, for a generalized dissipative coefficient, given in Eq.~(\ref{ups}), and any inflaton potential, $Q$ can be written in terms of $\phi$ as  \cite{Das:2020lut}
\begin{eqnarray}
(1+Q)^{2p}Q^{4-p}=\frac{M_{\rm Pl}^{2p+4}C_\Upsilon^4M^{4(1-p-c)}}{2^{2p}3^2C_R^p}\frac{V,_\phi^{2p}}{V^{p+2}}\phi^{4c}.\label{Q-phi}
\end{eqnarray}
If we consider the case studied in \cite{Arya:2017zlb}, we have $p=3$, $c=-2$ and $V=\lambda\phi^4$, yielding 
\begin{eqnarray}
(1+Q)^6Q=\frac{64}{9}\frac{C_\Upsilon^4\lambda}{C_R^3}\left(\frac{M_{\rm Pl}}{\phi}\right)^{10},
\end{eqnarray}
which can then be easily inverted to express $\phi$ as a function of $Q$ as \cite{Arya:2017zlb}
\begin{eqnarray}
\frac{\phi}{M_{\rm Pl}}=\left(\frac{64}{9}\frac{C_\Upsilon^4\lambda}{C_R^3}\frac{1}{Q(1+Q)^6}\right)^{1/10}.
\end{eqnarray}
On the other hand, for the scenario analyzed in \cite{Bastero-Gil:2017wwl}, $p=1$, $c=0$ and $V=\lambda\phi^4$, which then gives
\begin{eqnarray}
(1+Q)^2Q^3=\frac{4}{9}\frac{C_\Upsilon^4}{C_R\lambda}\left(\frac{M_{\rm Pl}}{\phi}\right)^{6},
\end{eqnarray}
which can also be easily inverted to yield  \cite{Bastero-Gil:2017wwl}
\begin{eqnarray}
\frac{\phi}{M_{\rm Pl}}=\left(\frac{4}{9}\frac{C_\Upsilon^4}{C_R\lambda}\frac{1}{Q^3(1+Q)^2}\right)^{1/6}.
\end{eqnarray}
However, this very last step where the expression relating $Q$ to $\phi$ is inverted to obtain $\phi$ as a function of $Q$ is very restrictive, and analytical results can only be obtained if one chooses to work with simpler inflaton potentials, like the quartic potential used in \cite{Bastero-Gil:2017wwl, Arya:2017zlb}. This very step, in general, is not achievable. Let's discuss an example. 

\begin{figure}[ht!]
\centering 
\includegraphics{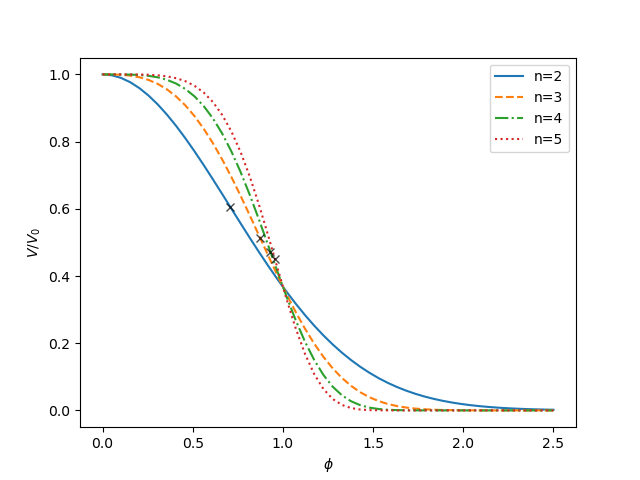}
\caption{\label{gen-exp-pot} The form of generalized exponential potential $V(\phi)=V_0e^{-\alpha(\phi/M_{\rm Pl})^n}$, where the $\phi$ values in the x-axis are given in units of $M_{\rm Pl}$.}
\end{figure}

Generalized exponential potentials of the form \cite{Geng:2015fla, Geng:2017mic, Ahmad:2017itq}
\begin{eqnarray}
V(\phi)=V_0e^{-\alpha(\phi/M_{\rm Pl})^n}, \label{exp-pot}
\end{eqnarray}
where $n$ is an integer, has the form given in Fig.~\ref{gen-exp-pot}. These potentials have an inflection point at 
\begin{eqnarray}
\frac{\phi_{\rm inflection}}{M_{\rm Pl}}=\left(\frac{n-1}{n\alpha}\right)^{1/n},
\end{eqnarray}
which have been shown with cross-marks in Fig.~\ref{gen-exp-pot}. Such a potential has a plateau region when $\phi\ll\phi_{\rm inflection}$. WI has been studied in this region in \cite{Lima:2019yyv} in weak dissipative regime where the form of the dissipative coefficient was taken as in Eq.~(\ref{cubic}).
However, one can see from Fig.~\ref{gen-exp-pot} that when $\phi>\phi_{\rm inflection}$ the potential becomes very steep. WI has been studied in these steep regions of the generalized exponential potentials in \cite{Das:2020xmh} in the strong dissipative regime where the dissipative coefficient was chosen of the form Eq.~(\ref{MWI}). We note that for such a non-trivial inflaton potential one can write $Q$ as a function of $\phi$, using Eq.~(\ref{Q-phi}), as 
\begin{eqnarray}
(1+Q)^6Q=\frac{\alpha^6n^6C_\Upsilon^4}{2^63^2C_R^3}\frac{M_{\rm Pl}^4V_0}{M^8}\left(\frac{\phi}{M_{\rm Pl}}\right)^{6n-6}e^{-\alpha(\phi/M_{\rm Pl})^n}.
\end{eqnarray}
This relation holds for dissipative coefficient given in Eq.~(\ref{MWI}), and a similar equation can also be written for the dissipative coefficient in Eq.~(\ref{cubic}). It is obvious that now it is not possible to invert the above equation to obtain an analytical form of $\phi$ as a function of $Q$. Therefore, the method illustrated in \cite{Bastero-Gil:2017wwl, Arya:2017zlb} to incorporate the scalar power spectrum of WI into CAMB fails in such cases.

\section{A generalized method to mould the WI power spectra in CAMB-ready forms}
\label{method}

A good look at the WI scalar (Eq.~(\ref{scalar-ps})) and tensor (Eq.~(\ref{tensor-ps})) power spectra would tell us that these power spectra depend only on $H$, $\dot\phi$, $T$ and $Q$: parameters whose evolutions are determined entirely by the background dynamics of WI, given by Eq.~(\ref{KG-eq}), Eq.~(\ref{rhor-eq}) and Eq.~(\ref{first-friedmann}). Once we write Eq.~(\ref{rhor-eq}) in terms of $T$ using Eq.~(\ref{rho-T}), we can see that the evolution equations of $\phi$ and $T$ are coupled. For any general dissipative coefficient $\Upsilon(\phi,T)$  these coupled equations can be written in terms of the $e-$foldings $N$ as 
\begin{eqnarray}
&&H(N)^2\phi''(N)+\left(3H(N)^2+\dot H(N)+\Upsilon(\phi,T)H(N)\right)\phi'(N)+V,_\phi(\phi(N))=0, \nonumber\\
&& T'(N)+T(N)=\frac{1}{4C_R}\Upsilon(\phi,T)T^{-3}(N)H(N)(\phi'(N))^2,
\end{eqnarray}
and for specific dissipative coefficients of the form given in Eq.~(\ref{ups}) the above equations become
\begin{eqnarray}
&&H(N)^2\phi''(N)+\left(3H(N)^2+\dot H(N)+C_\Upsilon M^{1-p-c}T^p(N)\phi^c(N)H(N)\right)\phi'(N)+V,_\phi(\phi(N))=0, \nonumber\\
&& T'(N)+T(N)=\frac{C_\Upsilon M^{1-p-c}}{4C_R}T^{p-3}(N)\phi^c(N)H(N)(\phi'(N))^2.
\label{coupled-eqs}
\end{eqnarray}
In the above set of equations $H(N)$ and $\dot H(N)$ can be written in terms of $\phi(N)$, $\phi'(N)$ and $T(N)$, using the Friedmann equations given in Eqs.~(\ref{first-friedmann}--\ref{second-friedmann}), as 
\begin{eqnarray}
H(N)&=&\left(\frac{V(\phi(N))+C_RT^4(N)}{3-\frac12(\phi'(N))^2}\right)^{1/2}, \label{HN}\\
\dot H(N)&=&-\frac12\left(H^2(\phi'(N))^2+\frac43C_RT^4(N)\right).
\end{eqnarray}
The $\phi$ equation of the coupled equations in Eq.~(\ref{coupled-eqs}) is a second order differential equation whereas the $T$ equation is a first order differential equation. Hence, three initial conditions, $\phi(N=0)$, $\phi'(N=0)$ and $T(N=0)$, are required to be set for numerically evolving these equations. 
After the evolution one would get $\phi$, $\phi'$ and $T$ as functions of $N$.\footnote{We have used the Python packages NumPy \cite{harris2020array} and SciPy \cite{Virtanen:2019joe} to numerically solve the background equations, and Matplotlib \cite{Hunter:2007} for plotting.} Therefore, having $\phi(N)$, $\phi'(N)$ and $T(N)$, all the required four quantities $\dot\phi(=H\phi')$, $H$ (as given in Eq.~(\ref{HN})), $T$ and $Q(=\Upsilon/3H)$ can be obtained as functions of the $e-$folds $N$. Thus, after employing this prescription one can obtain the primordial power spectra, both the scalar and the tensor, as a function of $N$. The end of inflation is marked when $\epsilon_H=-\dot H/H^2=1$.

The only task remains is to relate $N$ with $k$. For that we can relate the two comoving Hubble scales, the scale when the comoving wavenumber $k_*$ crosses the horizon during inflation $(k_*=a_*H_*)$, and the present Hubble scale $a_0H_0$ as  \cite{Liddle:2003as}
\begin{eqnarray}
\frac{k_*}{a_0H_0}=\frac{a_*}{a_{\rm end}}\frac{a_{\rm end}}{a_{\rm reh}}\frac{a_{\rm reh}}{a_0}\frac{H_*}{H_0}, \label{k-N}
\end{eqnarray}
where $a_{\rm end}$ and $a_{\rm reh}$ signify the scale factors at the end of inflation and the end of reheating, respectively, and we set the current scale factor $a_0=1$. We will show that in the cases which we will discuss, WI smoothly transits to a radiation dominated epoch, and no reheating takes place.\footnote{In some WI models, the Universe goes through a brief kination dominated epoch before entering into a radiation dominated epoch. In such cases, one needs to take into account the number of $e-$foldings spent in the kination period.} Thus, we can set $a_{\rm end}=a_{\rm reh}$. One can relate $a_{\rm end}$ with $a_0$ by entropy conservation as 
\begin{eqnarray}
g_s(T_{\rm end})T_{\rm{end}}^3a_{\rm end}^3=\left(2T_0^3+\frac{21}{4}T_{\nu,0}^3\right)a_0^3,
\end{eqnarray}
where $g_s(T_{\rm end})$ is the effective number of degrees of freedom at the end of WI, $T_0$ and $T_{\nu,0}=(4/11)^{1/3}T_0$ are the present day temperatures of the (CMB) photons and the cosmic neutrinos. Also, using $dN=d\ln a$ we see that $a_*/a_{\rm end}=e^{N_*-N_{\rm end}}$. Putting all these information together we can write Eq.~(\ref{k-N}) as 
\begin{eqnarray}
k_*=e^{N_*-N_{\rm end}} \left(\frac{43}{11g_s(T_{\rm end})}\right)^{1/3}\frac{T_0}{T_{\rm end}}H_*.
\end{eqnarray}
The values of $N_{\rm end}$ and $T_{\rm end}$ can be obtained from the evolution equation when $\epsilon_H=1$. We set $g_s(T_{\rm end})=106.75$ for definiteness, and $T_0=2.725$ K $=2.349\times 10^{-13}$ GeV. Using the above equation one can then numerically find the $e-$fold number $N_P$ when the pivot scale $k_P=0.05$ Mpc$^{-1}$ leaves the horizon during inflation as 
\begin{eqnarray}
e^{N_P}H_P=k_Pe^{N_{\rm end}} \left(\frac{43}{11g_s(T_{\rm end})}\right)^{-1/3}\frac{T_{\rm end}}{T_0}. \label{NP}
\end{eqnarray}
Once $N_P$ is known, one can integrate Eq.~(\ref{dkdN}) from $N=0$ to $N=N_P$ and then from $N=N_P$ to $N=N_{\rm end}$ to relate values of $k$ to that of $N$. Previously we have all the numerical values of the power spectra as a function of $N$, and now we have related $N$ values with $k$ values. Hence, by one-to-one mapping we now have the power spectra as functions of $k$, and these numerical values can be easily incorporated in CAMB. 

This prescribed methodology has certain advantages over the ones employed before in \cite{Bastero-Gil:2017wwl, Arya:2017zlb}, such as:
\begin{itemize}
\item The most important advantage of this methodology is that it is applicable to any form of inflaton potential, whereas the previous methods of \cite{Bastero-Gil:2017wwl, Arya:2017zlb} are applicable to only very simple forms of inflaton potentials, such as the quartic self-potential, and they fail when more complex inflaton potentials (such as the generalized exponential potential in Eq.~(\ref{exp-pot})) are considered in WI. 
\item The methodology adopted in \cite{Bastero-Gil:2017wwl, Arya:2017zlb} relies on slow-roll approximated dynamics of WI (for example, Eq.~(\ref{dQdN}) and Eq.~(\ref{Q-phi})). However, the one described here makes use of the full background dynamics of WI without any slow-roll approximation. Therefore, the power spectra (as a function of $k$) obtained using this methodology are more accurate than the ones obtained after the methods prescribed in  \cite{Bastero-Gil:2017wwl, Arya:2017zlb}. Moreover, as the numerical evolution of the background dynamics of this methodology doesn't require any slow-roll approximations,  it can easily be generalised for WI scenarios where the dynamics depart from slow-roll, e.g. ultraslow-roll \cite{Biswas:2023jcd} or constant-roll \cite{Biswas:2024oje}, though the power spectra for beyond slow-roll WI scenarios are yet to be developed \cite{Biswas:2023jcd, Biswas:2024oje}.
\item This methodology can be employed for any dissipative coefficient including those which cannot be written in the general form as in Eq.~(\ref{ups}), e.g. the one in Eq.~(\ref{EFT}). The previous methods of \cite{Bastero-Gil:2017wwl, Arya:2017zlb} are restricted only to WI models where the form of the dissipative coefficient can be written in the form as in Eq.~(\ref{ups}).
\end{itemize}

We note here that the numerical method based on Fokker-Planck approach to determine the WI scalar power spectrum developed in \cite{Ballesteros:2023dno} also makes use of full numerical evolutions of the background as well as the perturbations equations. Hence, in principle, this method, too, is not limited to slow-roll approximations of the background dynamics, and is potentially promising to be applicable to WI models with non-trivial potentials and generic dissipative coefficients. Yet, only simple WI models with quadratic, quartic and sextic potentials in combination with either of the three simple dissipative coefficients, $\Upsilon_{\rm linear}$, $\Upsilon_{\rm cubic}$ or $\Upsilon_{\rm MWI}$, have been explored in \cite{Ballesteros:2023dno}. It is thus not clear how well this newly developed technique will work for WI models with non-trivial dissipative coefficients and inflaton potentials. Moreover, as the method doesn't yield any analytic form of the scalar power spectrum to be fed into CAMB, the parameter space is scanned using only the best fit values of two cosmological parameters, namely $A_s$ and $n_s$ (only at the pivot scale) of the Planck observations. It is also not clear how the whole CMB TT spectrum can be compared with the numerical power spectrum generated by the method developed in \cite{Ballesteros:2023dno}. On the other hand, the MCMC analysis yield the best-fit model parameters of any inflationary model by comparing the full CMB data set (both temperature and polarization) of Planck observations while varying all the relevant cosmological parameters ($\Omega_bh^2$, $\Omega_ch^2$, $\tau$, $A_s$, $n_s$ and $H_0$). The method developed in \cite{Ballesteros:2023dno} in that sense is limited as only two cosmological parameters ($A_s$ and $n_s$) can be used to restrict the parameter space. This limitation can only be overcome if the numerically generated power spectrum of  \cite{Ballesteros:2023dno} can be fed into CAMB and a full MCMC analysis is done in the standard way, which will help further validate the numerical method introduced in  \cite{Ballesteros:2023dno}.

\section{Testing specific models of WI against the Planck data employing the prescribed generalized method}
\label{tests}

In this section, we will put to test a few WI models against the Planck data using the generalized method designed in this paper to obtain primordial power spectra as functions of $k$. First, we will discuss a model with generalized exponential potential (Eq.~(\ref{exp-pot})) for which the previous methods described in \cite{Bastero-Gil:2017wwl, Arya:2017zlb} fail. Then, we will verify, employing the new methodology, the results obtained in \cite{Bastero-Gil:2017wwl, Arya:2017zlb}. 

\subsection{WI models with generalized exponential potentials and $\Upsilon_{\rm MWI}$ as dissipative coefficient}

We stated before that the generalized exponential potential of the form given in Eq.~(\ref{exp-pot}) was explored in WI in \cite{Lima:2019yyv, Das:2020xmh}. In the first paper \cite{Lima:2019yyv}, the authors studied the model with $\Upsilon_{\rm cubic}$ where WI takes place in the plateau region of the potential and in weak dissipative regime. In the second paper \cite{Das:2020xmh}, WI was studied in the steep part of the potential ($\phi>\phi_{\rm inflection}$) in the strong dissipative regime where the choice of the dissipative coefficient was $\Upsilon_{\rm MWI}$. We will examine the WI model presented in the second paper \cite{Das:2020xmh} with the current Planck data. 

In this analysis, we will consider four different cases with $n=2,\,3,\,4,\,5$ as has been studied in \cite{Das:2020xmh}. We let the model parameters $g_*$, $V_0$, $C_\Upsilon$ and $\alpha$ float while feeding the model into Cobaya. The priors set for these model parameters are given in Table~\ref{priors} which are in accordance with the choice of parameters considered in \cite{Das:2020xmh} (see Table~1 of \cite{Das:2020xmh}). To evolve the background dynamics according to Eqs.~(\ref{coupled-eqs}) for such models, one needs to choose the initial conditions $\phi_0$, $\phi'_0$ and $T_0$, which designates the values of these parameters at $N=0$. The choice of the initial conditions as well as the best-fit model parameters are given in Table~\ref{table-mwi-1}. One can note that the best-fit values of the model parameters obtained after the MCMC analysis through Cobaya match closely with the choice of parameters given in Table~1 of  \cite{Das:2020xmh}.

\begin{table} [h!]
\begin{center}
	\begin{tabular}{|c|c|c|c|c|c|c|c|c|}
		\hline
		$n$ & \multicolumn{2}{c|}{$g_*$} & \multicolumn{2}{c|}{$V_0$ (GeV$^4$)} & \multicolumn{2}{c|}{$C_\Upsilon$} & \multicolumn{2}{c|}{$\alpha$} \\
		
		\hline
		& min & max & min & max & min & max & min & max \\
		
		\hline
		$2$ & $3$ & $600$ & $10^{37}$ & $10^{40}$ & $10^{-12}$ & $10^{-8}$ & $7$ & $15$ \\
		
		\hline
		$3$ & $3$ & $600$ & $10^{37}$ & $10^{40}$ & $10^{-12}$ & $10^{-8}$ & $1$ & $5$ \\
		
		\hline
		$4$ & $3$ & $600$ & $10^{37}$ & $10^{40}$ & $10^{-12}$ & $10^{-8}$ & $0$ & $3$ \\
		
		\hline
		$5$ & $3$ & $600$ & $10^{37}$ & $10^{40}$ & $10^{-12}$ & $10^{-8}$ & $0$ & $2$ \\
		\hline
	\end{tabular}
	\caption{{\label{priors}}Priors for model parameters for MCMC run.}
	\end{center}
\end{table}

\begin{table}[h!]
 	\begin{tabular}{|c|c|c|c|c|c|c|c|c|}
 		\hline
		\quad & \multicolumn{3}{c}{Initial Conditions} & \multicolumn{5}{|c|}{Model Parameters} \\ \hline
 		$n$ & $\phi_0$ (GeV) & $\phi'_0$ (GeV) & $T_0$ (GeV) & $g_*$ & $V_0$ (GeV$^4$) & $C_\Upsilon$ & $M$ (GeV) & $\alpha$ \\ 
 		
 		\hline
 		& & & & &  & & &  \\
 		$2$ & $5.74 \times 10^{17}$ & $10^{-5}$ & $10^{-5}$ & $124.727$ & $5.63 \times 10^{38}$ & $5.12\times10^{-11}$ & $9.6 \times 10^{5}$ & $9.17$  \\
 		
 		\hline
 		& & & & &  & & &  \\
 		$3$ & $1.47 \times 10^{18}$ & $10^{-5}$ & $10^{-5}$ & $111.053$ & $9.08\times 10^{38}$ & $5.16\times10^{-11}$ & $9.6 \times 10^{5}$ & $3.06$   \\
 		
 		\hline
 		& & & & &  & & & \\
 		$4$ & $1.91 \times 10^{18}$ & $10^{-5}$ & $10^{-5}$ & $81.7598$ & $1.03\times 10^{38}$ & $9.32 \times10^{-11}$ & $9.6 \times 10^{5}$ & $1.95$   \\
 		
 		\hline
 		& & & & &  & & &  \\
 		$5$ & $3.27 \times 10^{18}$ & $10^{-5}$ & $10^{-5}$ & $373.185$ & $7.39\times 10^{38}$ & $1.75\times10^{-10}$ & $9.6 \times 10^{5}$ & $0.18$  \\
 		\hline
 	\end{tabular}
	\caption{\label{table-mwi-1} {Best-fit values of the model parameters and the choice of the initial conditions.}}
 	
 \end{table}
 
\begin{center}
\begin{figure*}[h!]
\begin{subfigure}[t]{0.5\textwidth}
\centering
\includegraphics[width=7.5cm]{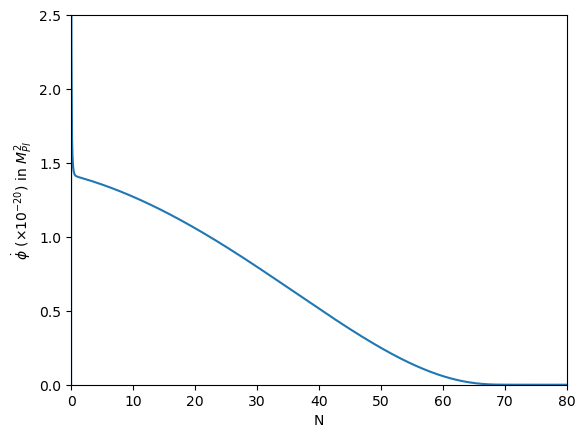}
\caption{Evolution of $\dot\phi$}
\end{subfigure}
\begin{subfigure}[t]{0.5\textwidth}
\centering
\includegraphics[width=7.5cm]{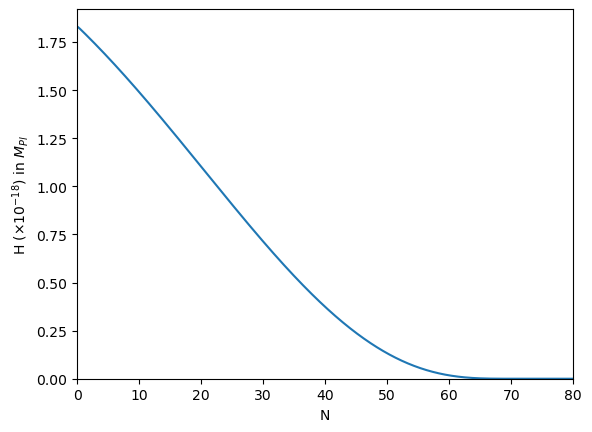}
\caption{Evolution of $H$}
\end{subfigure}\\
\begin{subfigure}[t]{0.5\textwidth}
\centering
\includegraphics[width=7.5cm]{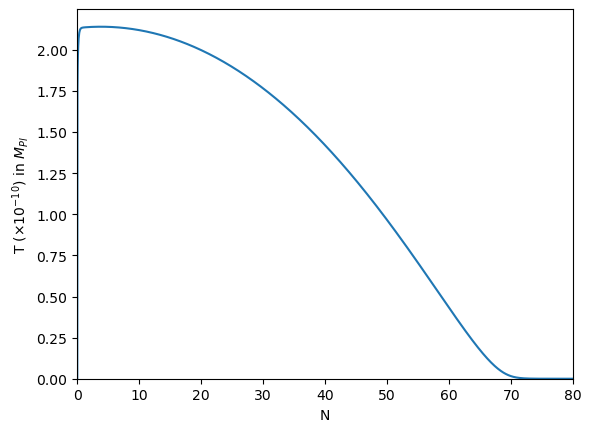}
\caption{Evolution of $T$}
\end{subfigure}
\begin{subfigure}[t]{0.5\textwidth}
\centering
\includegraphics[width=7.5cm]{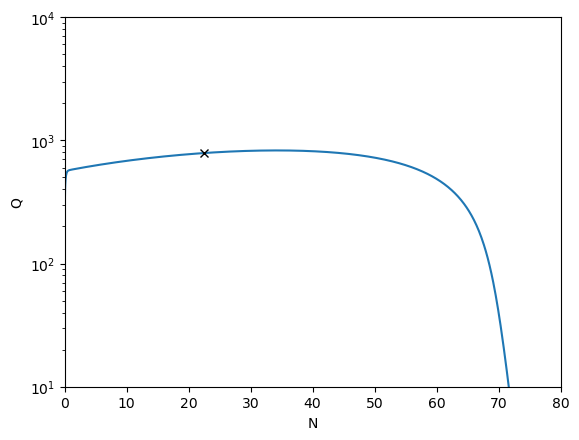}
\caption{Evolution of $Q$}
\end{subfigure}
\caption{Results of numerical evolution of the $n=2$ model.}
\label{n2-model}
\end{figure*}
\end{center}

\begin{center}
\begin{figure*}[h!]
\begin{subfigure}[t]{0.5\textwidth}
\centering
\includegraphics[width=7.5cm]{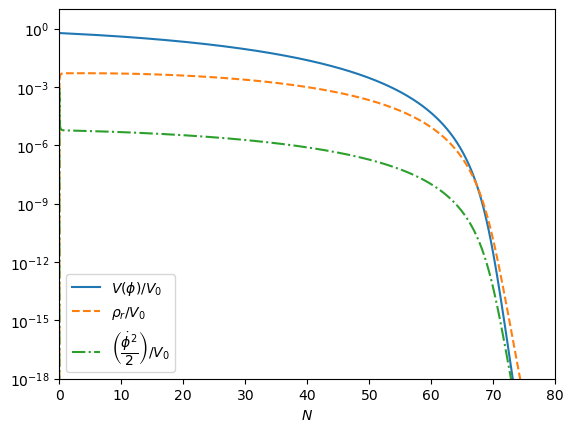}
\caption{Smooth transition to radiation dominated epoch after graceful exit}
\end{subfigure}
\begin{subfigure}[t]{0.5\textwidth}
\centering
\includegraphics[width=8.5cm]{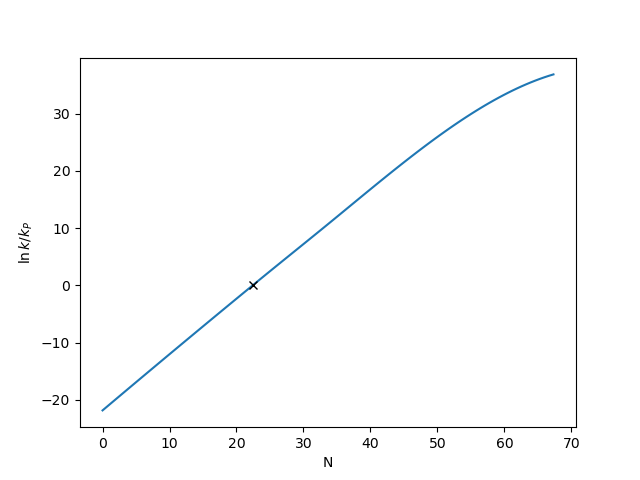}
\caption{$k$ vs. $N$ (the cross marks $k=k_P$)}
\end{subfigure}
\caption{The left panel shows that WI model with $n=2$ case smoothly transits to a radiation dominated epoch after graceful exit, whereas the right panel depicts $k$ as a function of $N$.}
\label{k-evolv}
\end{figure*}
\end{center}

To illustrate, we will consider the $n=2$ case. Similar results can be shown for $n=3,\,4,\,5$. The numerical values of the parameters required to determine the power spectra, $\dot\phi$, $H$, $T$, and $Q$, obtained after numerically evolving the coupled equations (Eqs.~(\ref{coupled-eqs})) are given in Fig.~\ref{n2-model}, where the best-fit model parameters furnished in Table~\ref{table-mwi-1} have been used. The cross-mark in the $Q$ vs $N$ plot indicates the value of $Q_*$, indicating that in this model of WI inflation takes place in the strong dissipative regime. We also note that the model smoothly transits to a radiation dominated era after graceful exit, as shown in the left panel of Fig.~\ref{k-evolv}, and we determine $N_{\rm end}-N_P=44.895$ using Eq.~(\ref{NP}). Hence employing the method discussed in the previous section, we determine $k$ as a function of $N$ as shown in the right panel of Fig.~\ref{k-evolv}.

\begin{figure}[h!]
\centering
\includegraphics[width=12.0cm]{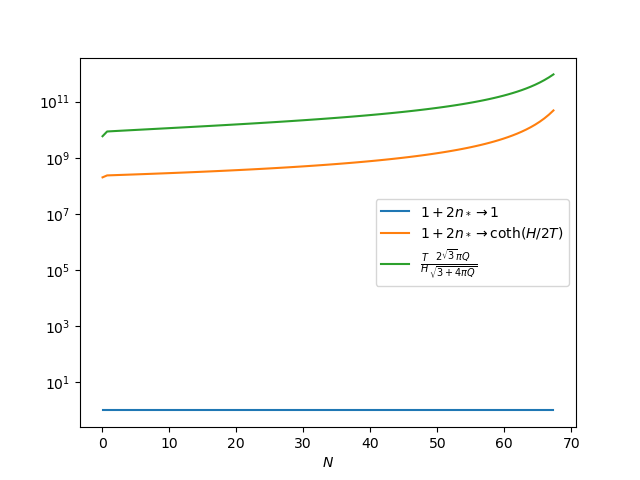}
\caption{To illustrate that the thermalization or non-thermalization of the inflaton field during WI has insignificant effect on the scalar power spectrum of WI}
\label{coth-plot}
\end{figure}

\begin{figure}[h!]
\centering
\includegraphics[width=12.0cm]{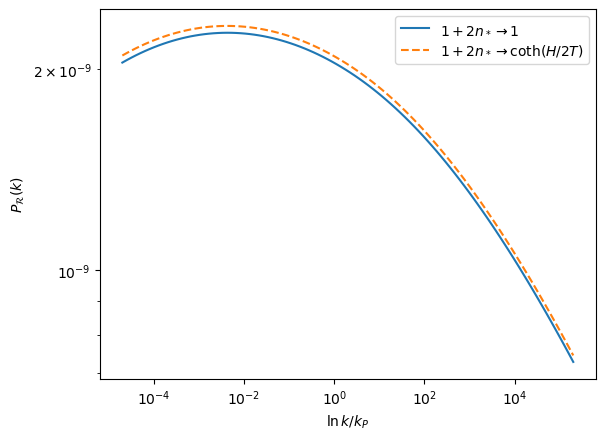}
\caption{The scalar power spectrum as a function of $k$ for the $n=2$ case. The blue solid line (orange dashed line) depicts the power spectrum when the inflaton doesn't (does) thermalize with the radiation bath.}
\label{ps-plot}
\end{figure}

The scalar power spectrum in WI can be determined in two cases, one when the inflaton doesn't thermalize with the radiation bath (thus by setting $n_*=0$ in Eq.~(\ref{scalar-ps})) and the other when the inflaton does thermalize with the radiation bath (thus by setting $1+2n_*=\coth(H_*/2T_*)$ in Eq.~(\ref{scalar-ps})). However, we note that in both these cases $1+2n_*$ remains subdominant by order(s) of magnitude w.r.t. the other term in the bracket in Eq.~(\ref{scalar-ps}), as can be seen from Fig.~\ref{coth-plot}. We plot in Fig.~\ref{ps-plot} the scalar power spectrum ${\mathcal P}_{\mathcal R}$ as a function of $k$ for both these cases to illustrate that the thermalization of the inflaton field has negligible effect of the scalar power spectrum in such a WI model. 

\begin{center}
\begin{figure*}[h!]
\begin{subfigure}[t]{0.5\textwidth}
\centering
\includegraphics[width=7.5cm]{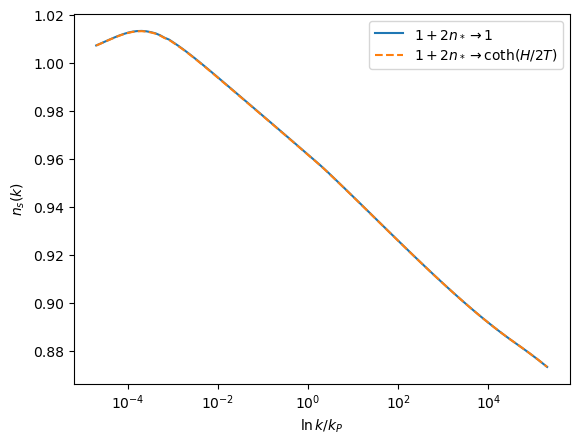}
\caption{$n_s$ as a function of $k$}
\end{subfigure}
\begin{subfigure}[t]{0.5\textwidth}
\centering
\includegraphics[width=7.5cm]{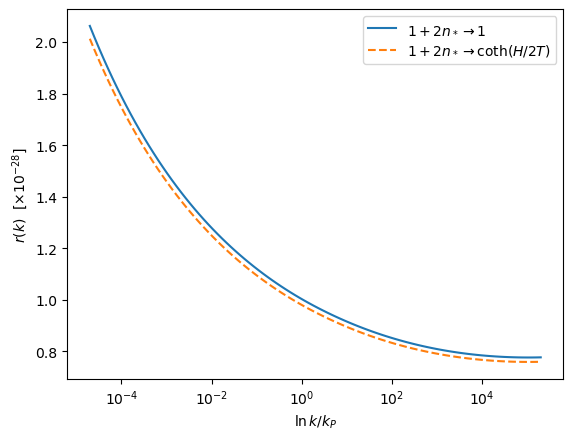}
\caption{$r$ as a function of $k$}
\end{subfigure}
\caption{$n_s$ and $r$ as a function of $k$ for the $n=2$ case. The blue solid line (orange dashed line) depicts $n_s$ and $r$ when the inflaton doesn't (does) thermalize with the radiation bath.}
\label{ns-r}
\end{figure*}
\end{center}

Once the scalar and tensor power spectra are known as functions of $k$, one can determine the scalar spectral index $n_s$ and the tensor-to-scalar ratio as 
\begin{eqnarray}
n_s-1=\frac{d\ln{\mathcal P}_{\mathcal R}(k)}{d\ln(k)}, \quad\quad \quad\quad r=\frac{{\mathcal P}_T}{{\mathcal P}_{\mathcal R}}.
\end{eqnarray}
Fig.~\ref{ns-r} shows both $n_s$ and $r$ as functions of $k$ for the $n=2$ case.

\begin{figure}[h!]
\centering
\includegraphics[width=16cm]{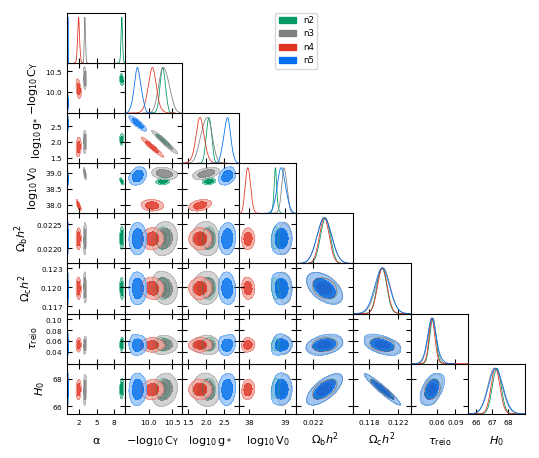}
\caption{The posterior distribution of all the four model parameters, $g_*$, $V_0$, $C_\Upsilon$ and $\alpha$ along with $\Lambda$CDM parameters, such as $\Omega_bh^2$ (baryon density parameter), $\Omega_ch^2$ (CDM density parameter), $\tau_{\rm reio}$ (optical depth) and $H_0$ (current Hubble parameter) for cases $n=2,\,3,\,4,\,5$.}
\label{contour-plot}
\end{figure}

\begin{figure}[h!]
\centering
\includegraphics[width=16cm]{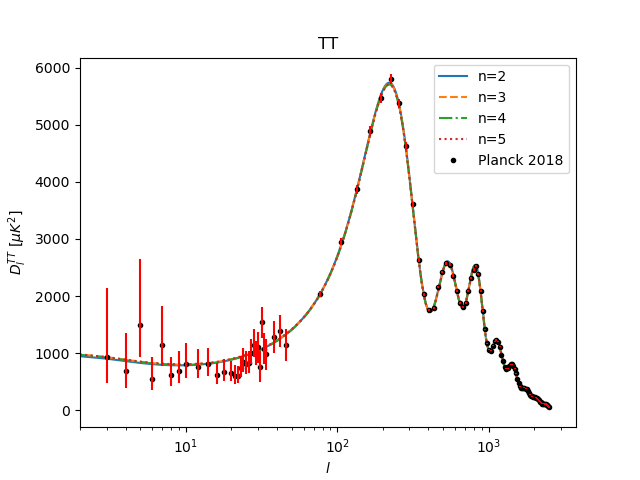}
\caption{The CMB TT spectrum for cases $n=2,\,3,\,4,\,5$ against Planck 2018 data.}
\label{TT-plot}
\end{figure}

Now coming back to all the models with $n=2,\,3,\,4,\,5$, we show the posterior distribution of all the four model parameters, $g_*$, $V_0$, $C_\Upsilon$ and $\alpha$ along with $\Lambda$CDM parameters, such as $\Omega_bh^2$ (baryon density parameter), $\Omega_ch^2$ (CDM density parameter), $\tau_{\rm reio}$ (reionization optical depth) and $H_0$ (current Hubble parameter), in Fig.~\ref{contour-plot}. The posterior values of the model parameters are given in Table~\ref{post-values} as well. The best-fit model parameters yield the scalar amplitude $(A_s)$, the scalar spectral index $(n_s)$ and the running of the scalar spectral index ($\alpha_s=dn_s/d\ln  k$) at pivot scale $k_P=0.05$ Mpc$^{-1}$ and $r$ at $k_P=0.002$ Mpc$^{-1}$ as given in Table~\ref{spectral}. 
Note that, in \cite{Das:2020xmh}, $r$ has been determined at $k_P=0.05$ Mpc$^{-1}$, and hence the values of $r$ obtained in \cite{Das:2020xmh} are slightly off than the ones obtained here. Moreover, values of $\alpha_s$ for these models have been studied in \cite{Das:2022ubr}, and the values of $\alpha_s$ quoted in Table IV of \cite{Das:2022ubr} match reasonably well with the ones obtained here. Above all, the scalar power spectrum obtained from the best-fit model parameters fit the Planck data very well as depicted in Fig.~\ref{TT-plot}. Note that all these results are quoted for $n_*\neq0$ case, i.e., for the case when the inflaton thermalizes with the radiation bath. 

\begin{table}[h!]
\begin{center}
 	\begin{tabular}{|c|c|c|c|c|}
 		\hline
 		$n$ & $g_*$ & $V_0$ (GeV$^4)$ & $C_\Upsilon$ & $\alpha$  \\ 
 		
 		\hline
 		&  &  & &    \\
 		$2$ & $121.62\pm 19.83$ & $(5.45 \pm 0.54) \times 10^{38}$ & $(5.07 \pm 0.65)\times10^{-11}$ & $9.22 \pm 0.13$   \\
 		
 		\hline
 		&  &  & &   \\
 		$3$ & $106.38\pm 33.91$ & $(9.69\pm 1.65)\times 10^{38}$ & $(4.82\pm 1.14)\times10^{-11}$ & $3.029\pm 0.098$  \\
 		\hline
 		
 		\hline
 		&  &  & &    \\
 		$4$ & $70.86 \pm  20.58$ & $(9.53\pm 1.60)\times 10^{37}$ & $(8.61 \pm 1.85) \times10^{-11}$ & $1.98 \pm 0.17$   \\
 		\hline
 		
 		\hline
 		&  &  & &    \\
 		$5$ & $393.52 \pm  87.57$ & $(8.24 \pm 2.25)\times 10^{38}$ & $(1.74\pm 0.30)\times10^{-10}$ & $0.188 \pm 0.026$  \\
 		\hline
 	\end{tabular}
 	\caption{{\label{post-values}}Posterior distribution of the model parameters.}
 \end{center}
 \end{table}
 
 \begin{table}[h!]
 \begin{center}
	\begin{tabular}{|c|c|c|c|c|}
		\hline
		 $n$ & $\ln (10^{10}A_s)$ & $n_s$ & $\alpha_s$ & $r_{0.002}$ \\
		 
		 \hline
		 &  &  & &    \\
		 $2$ & $3.0399$ & $0.9617$ & $-0.0071$ & $1.15 \times 10^{-28}$ \\
		 
		 \hline
		 &  &  & &    \\
		  $3$ & $3.0407$ & $0.9609$ & $-0.0049$ & $9.50\times 10^{-29}$ \\
		  
		 \hline
		 &  &  & &    \\
		 $4$ & $3.0396$ & $0.9638$ & $-0.0043$ & $1.34\times 10^{-29}$ \\
		 
		 \hline
		 &  &  & &    \\
		 $5$ & $3.0322$ & $0.9596$ & $-0.0039$ & $5.29\times 10^{-29}$ \\
		 \hline
	\end{tabular}\caption{{\label{spectral}}$A_s$, $n_s$, $\alpha_s$ and $r$ from the best-fit model parameters}
\end{center}
\end{table}

\subsection{WI model with quartic potential and $\Upsilon_{\rm linear}$ as dissipative coefficient}

WI model with quartic potential with $\Upsilon_{\rm linear}$ as dissipative coefficient has been studied in  \cite{Bastero-Gil:2017wwl}. The analysis was done with the then current data of Planck 2015 and BICEP2/Keck Array \cite{BICEP2:2015xme, Planck:2015sxf}. Moreover, in this paper, the number of $e-$folds is counted considering a reheating phase at the end of WI. 

\begin{center}
\begin{figure*}[h!]
\begin{subfigure}[t]{0.5\textwidth}
\centering
\includegraphics[width=7.5cm]{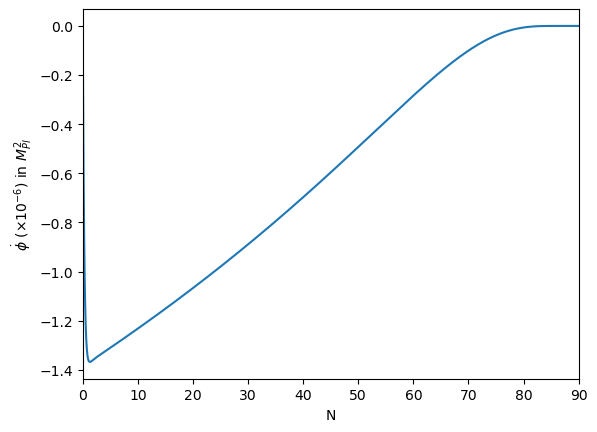}
\caption{Evolution of $\dot\phi$}
\end{subfigure}
\begin{subfigure}[t]{0.5\textwidth}
\centering
\includegraphics[width=7.5cm]{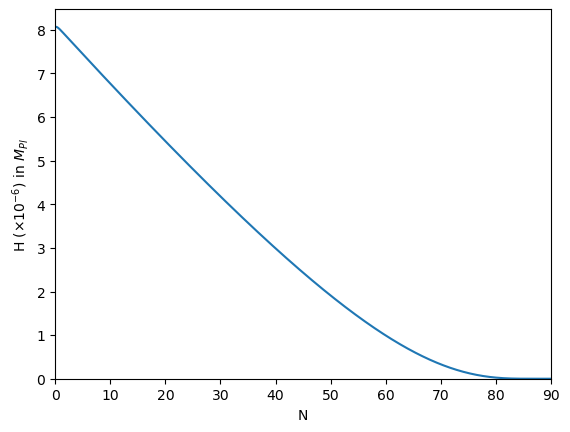}
\caption{Evolution of $H$}
\end{subfigure}\\
\begin{subfigure}[t]{0.5\textwidth}
\centering
\includegraphics[width=7.5cm]{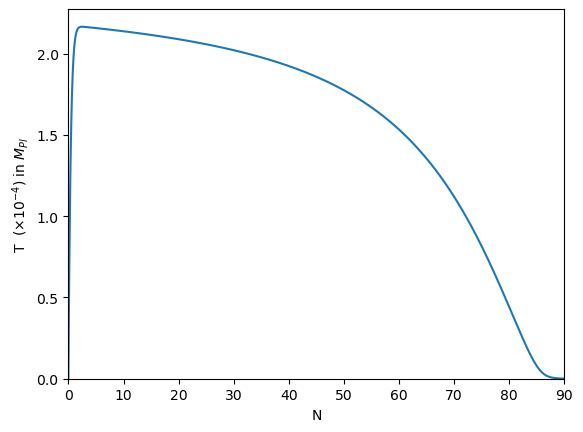}
\caption{Evolution of $T$}
\end{subfigure}
\begin{subfigure}[t]{0.5\textwidth}
\centering
\includegraphics[width=7.5cm]{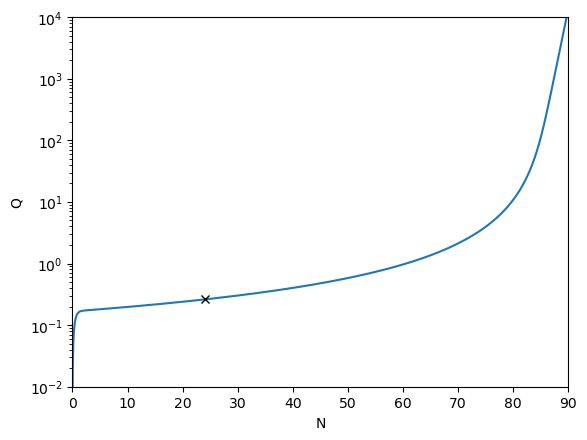}
\caption{Evolution of $Q$}
\end{subfigure}
\caption{Results of numerical evolution of the model studied in \cite{Bastero-Gil:2017wwl}.}
\label{MB-model}
\end{figure*}
\end{center}

\begin{center}
\begin{figure*}[h!]
\begin{subfigure}[t]{0.5\textwidth}
\centering
\includegraphics[width=7.5cm]{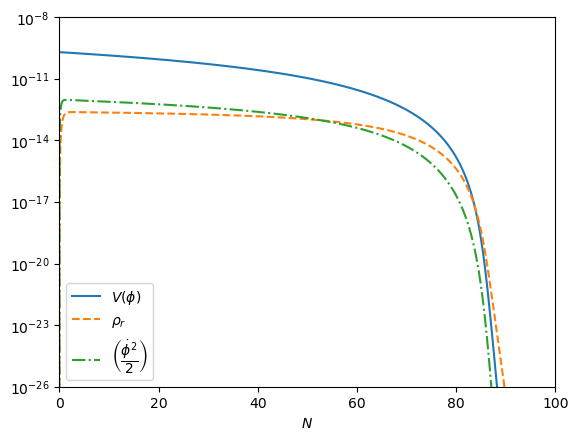}
\caption{Smooth transition to radiation domination}
\end{subfigure}
\begin{subfigure}[t]{0.5\textwidth}
\centering
\includegraphics[width=7.5cm]{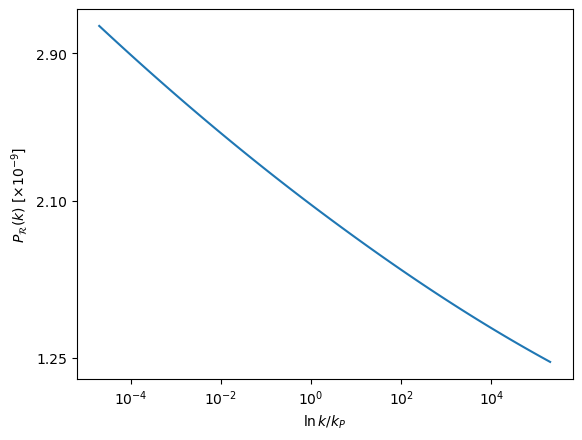}
\caption{Scalar power spectrum as a function of $k$}
\end{subfigure}
\caption{The left panel shows that the model studied in \cite{Bastero-Gil:2017wwl} smoothly transits into a radiation dominated epoch. The right panel plots the scalar power spectrum ${\mathcal P}_{\mathcal R}$ as a function of $k$.}
\label{ED-ps-MB}
\end{figure*}
\end{center}

After evolving the background equations given in Eqs.~(\ref{coupled-eqs}) for this WI model, we obtained the evolution of the required parameters as functions of $N$ as shown in Fig.~\ref{MB-model}. Here we have used the best-fit model parameters given in Table~\ref{MB-table}, and the cross-mark in the $Q$ vs $N$ plot indicates the value of $Q_*$ as before. We note that in this WI model inflation takes place in the weak dissipative regime ($Q_*<1$). We also see from the left panel of Fig.~\ref{ED-ps-MB} that this model of WI, too, smoothly transits into a radiation dominated epoch. Thus, our method to determine $k$ as a function of $N$ works in this case, and we determine $N_{\rm end}-N_P=59.58$. The scalar power spectrum is shown in the right panel of Fig.~\ref{ED-ps-MB} as a function of $k$. After the MCMC analysis of the model (with $n_*\neq0$) against the Planck 2018 data, we show the posterior distribution of the model parameters, $g_*$, $\lambda$ and $C_{\Upsilon}$, and the $\Lambda$CDM parameters, $\Omega_bh^2$, $\Omega_ch^2$, $\tau_{\rm reio}$ and $H_0$, in Fig.~\ref{contour-MB} and Table~\ref{MB-table}. Comparing Table~\ref{MB-table} with Table~1 of \cite{Bastero-Gil:2017wwl}, one can see that both the analysis produce nearly the same best-fit model parameters. This validates the  methodology prescribed in this paper. 

\begin{figure}[h!]
\centering
\includegraphics[width=12.0cm]{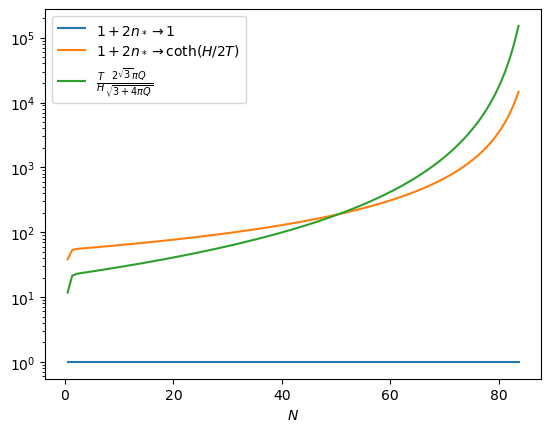}
\caption{To illustrate that the thermalization of the inflaton field during WI has significant effect on the scalar power spectrum of WI in the model with quartic potential and $\Upsilon_{\rm linear}$.}
\label{cothMB-plot}
\end{figure}

\begin{figure}[h!]
\centering
\includegraphics[width=15cm]{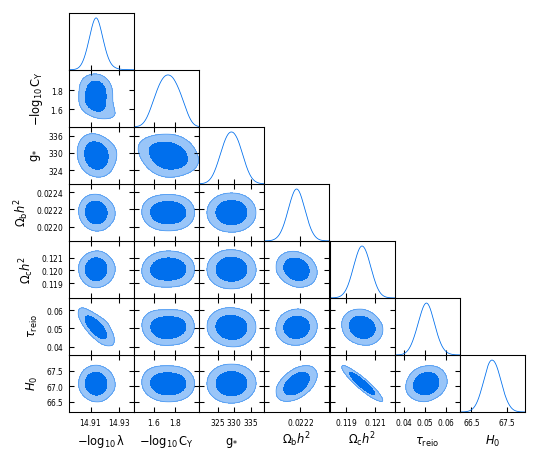}
\caption{The posterior distribution of model parameters, $g_*$, $\lambda$ and $C_\Upsilon$  along with $\Lambda$CDM parameters, $\Omega_bh^2$, $\Omega_ch^2$, $\tau_{\rm reio}$ and $H_0$ for  WI model studied in \cite{Bastero-Gil:2017wwl}}
\label{contour-MB}
\end{figure}

It is important to note that unlike the previous case, it becomes crucial in this case whether one considers the thermalization of the inflaton field while determining the form of the scalar power spectrum according to Eq.~(\ref{scalar-ps}). As one can see from Fig.~\ref{cothMB-plot}, in this case the factor $1+2n_*$ dominates over the other term in the bracket of Eq.~(\ref{scalar-ps}) when the pivot scale leaves the horizon. This issue has already been noticed in \cite{Bastero-Gil:2017wwl} (see Fig.~1 of \cite{Bastero-Gil:2017wwl}). Following the arguments provided in Sec.~4.5 of \cite{Ballesteros:2023dno}, one notes that in this model $Q\propto T/H$. Thus, in the strong dissipative regime, i.e. when  $Q_*\gg1$, $n_*$ can become relevant, however it remains subdominant to the other term in the bracket of Eq.~(\ref{scalar-ps}) as the other term is proportional to $Q_*^{3/2}$ (this is the reason of $1+2n_*$ being ineffective in the previous case). However, in the weak dissipative regime, $Q_*<1$,  $1+2n_*$ becomes relevant. This is exactly what is happening in this case as WI is taking place mostly in a weak dissipative regime as can be seen from Fig.~\ref{MB-model}(d).

\begin{table}[h!]
 \begin{center}
 	\begin{tabular}{|c|c|}
 		\hline
 		Parameter & $n_* \neq 0$    \\ 
 		
 		\hline
 		
 		$\Omega_b h^2$ & $0.02216 \pm 0.00008$    \\
 		
 		\hline
 		
 		$\Omega_c h^2$ & $0.1201\pm 0.0006$  \\
 		\hline
 		
 		\hline

 		\hline
 		
 		$-\log_{10} \lambda$ & $14.914 \pm  0.006$    \\
 		\hline
 		
 		$\lambda$ & $1.219 \times 10^{-15}$    \\
 		\hline
 		
 		\hline
 		
 		$-\log_{10} C_\Upsilon$ & $1.725\pm  0.109$    \\
 		\hline
 		
 		$C_\Upsilon$ & $0.0188$    \\
 		\hline
 		
 		\hline
 		
 		$g_*$ & $330.917\pm  2.978$    \\
 		\hline
 		
 		\hline
 		
 		$\tau$ & $0.0502\pm  0.0041$    \\
 		
 		\hline
 	\end{tabular}
 	\caption{{\label{MB-table}} Posterior distribution of the model parameters of the WI model studied in \cite{Bastero-Gil:2017wwl}.}
 \end{center}
 \end{table}

\subsection{WI model with quartic potential and $\Upsilon_{\rm cubic}$ as dissipative coefficient}

 We will reverify the WI model having $\Upsilon_{\rm cubic}$ as dissipative coefficient with quartic potential as has been studied in \cite{Arya:2017zlb}. In \cite{Arya:2017zlb}, the WI model has been constrained using the then available Planck 2015 data \cite{Planck:2015sxf}. The authors keep $N_{\rm{end}}-N_P$ as a free parameter and chose two cases with $N_{\rm{end}}-N_P=50$ or 60. They also keep the value of $Q$ at the pivot scale as a free parameter. 

\begin{center}
\begin{figure*}[h!]
\begin{subfigure}[t]{0.5\textwidth}
\centering
\includegraphics[width=7.5cm]{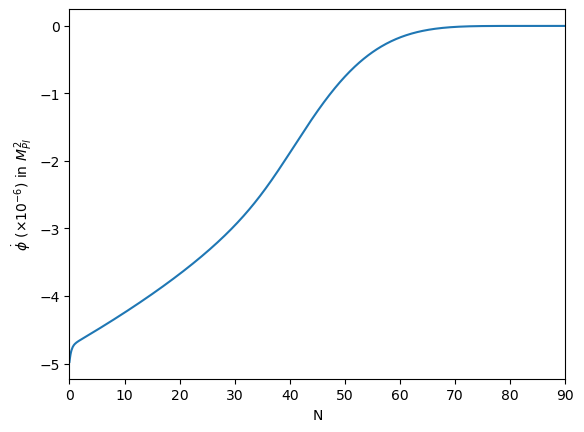}
\caption{Evolution of $\dot\phi$}
\end{subfigure}
\begin{subfigure}[t]{0.5\textwidth}
\centering
\includegraphics[width=7.5cm]{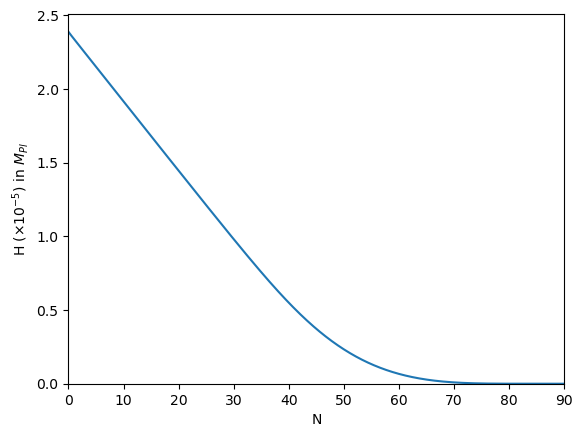}
\caption{Evolution of $H$}
\end{subfigure}\\
\begin{subfigure}[t]{0.5\textwidth}
\centering
\includegraphics[width=7.5cm]{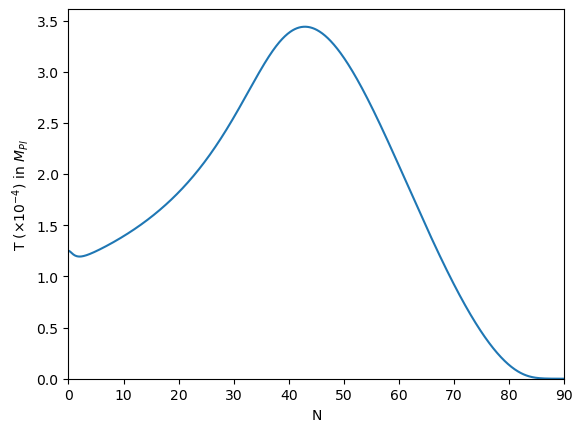}
\caption{Evolution of $T$}
\end{subfigure}
\begin{subfigure}[t]{0.5\textwidth}
\centering
\includegraphics[width=7.5cm]{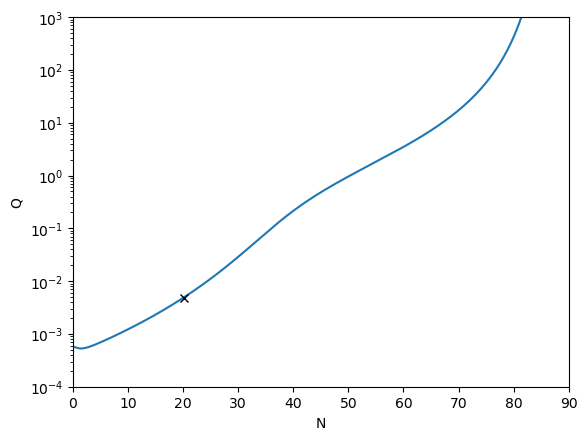}
\caption{Evolution of $Q$}
\end{subfigure}
\caption{Results of numerical evolution of the model studied in  \cite{Arya:2017zlb}.}
\label{RA-model}
\end{figure*}
\end{center}

\begin{center}
\begin{figure*}[h!]
\begin{subfigure}[t]{0.5\textwidth}
\centering
\includegraphics[width=7.5cm]{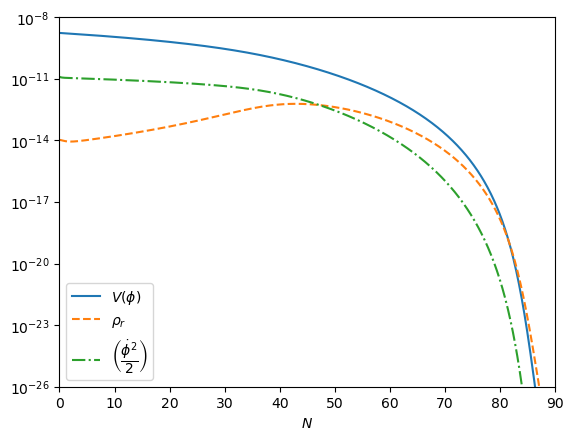}
\caption{Smooth transition to radiation domination}
\end{subfigure}
\begin{subfigure}[t]{0.5\textwidth}
\centering
\includegraphics[width=7.5cm]{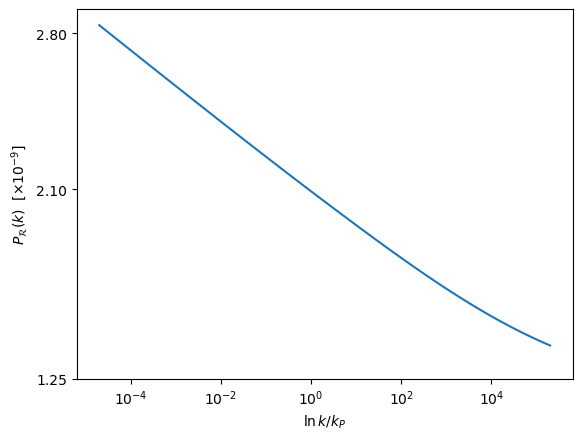}
\caption{Scalar power spectrum as a function of $k$}
\end{subfigure}
\caption{The left panel shows that the model studied in  \cite{Arya:2017zlb} smoothly transits into a radiation dominated epoch. The right panel plots the scalar power spectrum ${\mathcal P}_{\mathcal R}$ as a function of $k$.}
\label{ED-ps-RA}
\end{figure*}
\end{center}

\begin{figure}[h!]
\centering
\includegraphics[width=15cm]{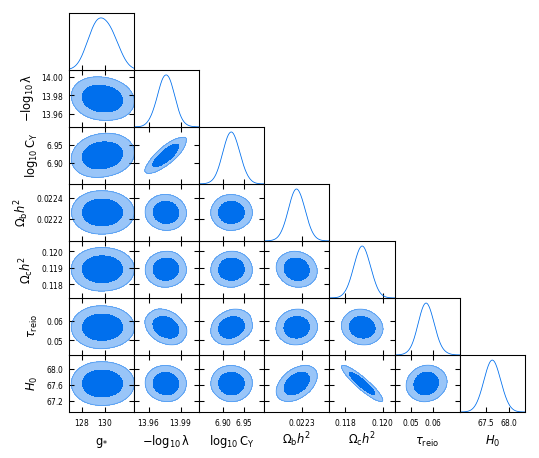}
\caption{The posterior distribution of model parameters, $g_*$, $\lambda$ and $C_\Upsilon$  along with $\Lambda$CDM parameters, $\Omega_bh^2$, $\Omega_ch^2$, $\tau_{\rm reio}$ and $H_0$ for  WI model studied in  \cite{Arya:2017zlb}.}
\label{contour-RA}
\end{figure}

\begin{figure}[h!]
\centering
\includegraphics[width=12.0cm]{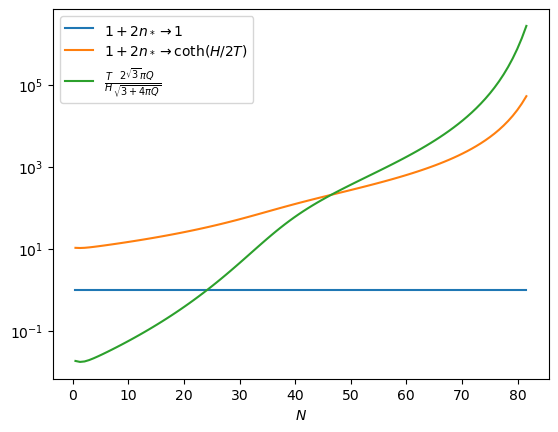}
\caption{To illustrate that the thermalization of the inflaton field during WI has significant effect on the scalar power spectrum of WI in the model with quartic potential and $\Upsilon_{\rm cubic}$.}
\label{cothRA-plot}
\end{figure}

In our prescribed methodology, $Q$ at the pivot scale is no longer a free parameter. The evolution of the required parameters as functions of $N$ are shown in Fig.~\ref{RA-model}, which have been plotted with the best-fit values of the model parameters furnished in Table~\ref{RA-table}. The cross-mark in Fig.~\ref{RA-model}(d) represents $Q_*$ which signifies that in this model WI has taken place in the weak dissipative regime. We note from the left panel of Fig.~\ref{ED-ps-RA} that this WI model smoothly transits into a radiation dominated epoch post inflation, and thus one can use our methodology to relate $k$ with $N$, and we obtain $N_{\rm end}-N_p=61.46$ for this model. The scalar power spectrum for this model is shown as a function of $k$ in the right panel of Fig.~\ref{ED-ps-RA}. After the MCMC analysis of the model (with $n_*\neq0$) against the Planck 2018 data, we show the posterior distribution of the model parameters, $g_*$, $\lambda$ and $C_{\Upsilon}$, along with the $\Lambda$CDM parameters, $\Omega_bh^2$, $\Omega_ch^2$, $\tau_{\rm reio}$ and $H_0$, in Fig.~\ref{contour-RA} and Table~\ref{RA-table}. Comparing  Table~\ref{RA-table} with Table~1 of \cite{Arya:2017zlb} (the column for $N_P=60$) we note that the posterior parameters match reasonably well. This again validates the methodology prescribed in this paper. Above all, like the previous case, in this case too the scalar power spectrum will crucially depend on whether one considers thermalization of the inflaton field during WI. As one can read from Fig.~\ref{RA-model}(d) that $Q_*<1$, and thus $1+2n_*$ contributes significantly to the scalar power spectrum as can be seen from Fig.~\ref{cothRA-plot}.

\begin{table}[h!]
\begin{center}
 	\begin{tabular}{|c|c|}
 		\hline
 		Parameter & $n_{*} \neq 0$    \\ 
 		
 		\hline

 		$\Omega_b h^2$ & $0.02226 \pm 0.00007$    \\
 		
 		\hline

 		$\Omega_c h^2$ & $0.1189\pm 0.0004$  \\
 		\hline
 		
 		\hline

 		\hline

 		$-\log_{10} \lambda$ & $13.976 \pm  0.008$    \\
 		\hline
 		
 		$\lambda$ & $1.057 \times10^{-14}$    \\
 		\hline
 		
 		\hline
 		
 		$\log_{10} C_\Upsilon$ & $6.9203\pm  0.0208$    \\
 		\hline
 		
 		$C_\Upsilon$ & $8.323 \times 10^6$    \\
 		\hline
 		
 		\hline
 		
 		$g_*$ & $129.638\pm  0.868$    \\
 		\hline
 		
 		\hline
 		
 		$\tau$ & $0.0569\pm  0.0038$    \\
 		
 		\hline

 	\end{tabular}
 	\caption{{\label{RA-table}} Posterior distribution of the model parameters of the WI model studied in \cite{Arya:2017zlb}.}
 \end{center}
\end{table}

\section{Conclusion}
\label{conclusion}

In this paper, we have devised a generalized methodology to incorporate the WI primordial power spectra, both scalar and tensor, in CAMB \cite{Lewis:1999bs, Howlett:2012mh} which is essential to perform a MCMC analysis of such WI models given the present cosmological data employing the publicly available codes, like CosmoMC  \cite{Lewis:2002ah, Lewis:2013hha} or Cobaya \cite{Torrado:2020dgo}. The prescribed method is applicable to all WI models with any kind of dissipative coefficient and potential. Previous methods, as in \cite{Bastero-Gil:2017wwl, Arya:2017zlb}, were rather restrictive in the sense that they were only applicable to WI models with specific forms of dissipative coefficients, as in Eq.~(\ref{ups}), and very simple forms of inflaton potentials, such as the quartic potential as has been exploited in \cite{Bastero-Gil:2017wwl, Arya:2017zlb}. Moreover, the method prescribed in this paper employs the full background dynamics of WI, rather than the slow-roll approximated ones as required by previous methods \cite{Bastero-Gil:2017wwl, Arya:2017zlb}, and thus, is generalized enough to be extended to beyond-slow-roll dynamics of WI, such as ultraslow-roll \cite{Biswas:2023jcd} or constant-roll \cite{Biswas:2024oje}. In addition, as the prescribed method doesn't call for slow-roll approximated dynamics, the primordial power spectra as functions of $k$ obtained from this mechanism are more accurate than the ones one can obtain by employing previous methodologies \cite{Bastero-Gil:2017wwl, Arya:2017zlb}. 

To illustrate our prescribed methodology, we analysed a WI model with generalized exponential potentials (Eq.~(\ref{exp-pot})), a case which is not possible to analyse by the previous methodologies \cite{Bastero-Gil:2017wwl, Arya:2017zlb} because of the complex form of such potentials. We chose $\Upsilon_{\rm MWI}$ as the dissipative coefficient and analyzed the model in the steep part of such potentials, as has been first analyzed in \cite{Das:2020xmh}. We did an MCMC analysis of the model parameters along with other $\Lambda$CDM parameters using the Planck 2018 data, and have shown that our analysis matches reasonably well with the parameter values chosen in \cite{Das:2020xmh}. We then employed our methodology to the models of those previous literature \cite{Bastero-Gil:2017wwl, Arya:2017zlb}, and showed that our results agree quite well with the results presented there. This exercise validates the functionality of our new methodology. Therefore, this new methodology that we prescribed in this paper will help put to test any model of WI against the concurrent cosmological data in future. 


\acknowledgments

The work of S.D. is supported by the Start-up Research Grant (SRG) awarded by Anusandhan National Research Foundation (ANRF), Department of Science and Technology, Government of India 
[File No. SRG/2023/000101/PMS]. S.D. is also thankful to Axis
Bank and acknowledges the financial support obtained
from them which partially supports this research. The authors acknowledge
the High Performance Computing (HPC) Facility of the
Ashoka University (The Chanakya @ Ashoka), used for
the MCMC runs. U.K. is  thankful to Gabriele Montefalcone 
and Alejandro Perez Rodriguez for useful discussions
regarding numerical computation of Warm Inflationary power spectrum. U.K. also acknowledges the
help extended by Dipankar Bhattacharya in the numerical
simulations. S.D.
thanks Rudnei Ramos for many useful discussions on
Warm Inflation.



\end{document}